\documentstyle[12pt]{article}
\input epsf

\newcommand{\sect}[1]{\setcounter{equation}{0}\section{#1}}

\newcommand\M{{\cal M}}

\newcommand\B{{\cal B}}
\newcommand\h{{\cal H}}
\newcommand\R{{\mathrm {I\!R}}}

\newcommand{\be}{\begin{equation}}
\newcommand{\ee}{\end{equation}}

\newcommand{\ssj}[6]{\left| \begin{array}{ccc} #1 & #2 & #3 \\ #4 & #5 & #6 \end{array} \right|}

\newcommand{\ra}{\rightarrow}

\newcommand{\lpq}{$L_{p,q}$}

\newcommand{\cqg}{Class. Quantum Grav.}

\newcommand{\jmp}{J. Math. Phys.}
\newcommand{\np}{Nucl. Phys.}

\newtheorem{lemma}{Lemma}

\begin{document}

\vspace{.5in}

\begin{titlepage}

\begin{flushright}
DAMTP-1998-65\\
June 1998\\
gr-qc/9806027
\end{flushright}

\vspace{.5in}

\begin{center}
{\LARGE \bf
Lens Spaces and Handlebodies in 3D Quantum Gravity}\\

\vspace{.4in}
{\large Radu Ionicioiu
     \footnote{\small\tt email:~R.Ionicioiu@damtp.cam.ac.uk}
     \footnote{\it on leave from Institute of Gravitation and Space Sciences, 21-25 Mendeleev Street, 70168 Bucharest, Romania} 
and Ruth M. Williams \footnote{\small\tt email:~R.M.Williams@damtp.cam.ac.uk}\\
~~\\
       {\small\it DAMTP, University of Cambridge}\\
       {\small\it Silver Street, Cambridge CB3 9EW, UK}\\}
\end{center} 

\vspace{.5in}
\begin{center}
{\large\bf Abstract}
\end{center}

\begin{center}
\begin{minipage}{4.75in}
{\small

We calculate partition functions for lens spaces $L_{p,q}$ up to $p=8$ and for genus 1 and 2 handlebodies $\h_1$, $\h_2$ in the Turaev-Viro framework. These can be interpreted as transition amplitudes in 3D quantum gravity. In the case of lens spaces \lpq\ these are vacuum-to-vacuum amplitudes $\O \ra \O$, whereas for the 1- and 2-handlebodies $\h_1$, $\h_2$ they represent genuinely topological transition amplitudes $\O \ra T^2$ and $\O \ra T^2 \# T^2$, respectively.

~~
}
\end{minipage}
\end{center}

\end{titlepage}
\addtocounter{footnote}{-1}

\sect{Introduction}

The search for a consistent theory of quantum gravity has been a major preocupation of many theoretical physicists for a long time, with the problems being both conceptual and technical. Although the goal is to find a theory appropriate to the apparently 4-dimensional world in which we live, considerably effort has gone into formulating theories in lower (and higher) dimensions. In particular, for three dimensions it is hoped that it might somehow be possible to generalize a theory to four dimensions, or at least gain some understanding of some of the conceptual issues involved.

In this paper we shall focus on the state sum invariants formulated by Turaev and Viro \cite{tv}. The link between the Turaev-Viro model and quantum gravity in 3 dimensions can be seen in two ways. Firstly, the TV partition function is the regularized version of the Ponzano-Regge model \cite{pr} of 3D quantum gravity. Secondly, the TV partition function is the square of that of a Chern-Simons model, which, in turn, is equivalent to 3D gravity, as shown by Witten \cite{witten} (see also \cite{jb} for the significance of TQFTs in quantum gravity and \cite{krasnov} for a link with $BF$ theory).

Turaev and Viro wrote down their state sum both for closed manifolds and for manifolds with boundary, which means that their expression can be interpreted as a transition amplitude for topology change between the boundaries involved or the vacuum, as appropriate.

The state sum model is for triangulated manifolds. The advantage of working in the P.L. category rather than the continuum is that there is a finite number of variables and the partition function is purely combinatorial. Unfortunately, this does not necessarily mean that it is easy to evaluate analytically, except in the simplest cases. For this reason, we make use of various methods of building three-manifolds from more elementary objects. In particular, it is known that any closed, orientable three-manifold can be constructed from two genus $g$ handlebodies by identifying their boundaries under a suitable homeomorphism $h: T_g \ra T_g$ (this is known as a {\it genus-g Heegaard decomposition/splitting}).

The idea is to obtain general expressions for the TV state sum for genus $g$ handlebodies and a prescription for gluing them together, either over their whole boundaries to construct closed manifolds, or over part of their boundaries to form higher genus handlebodies with which to construct other manifolds.

In the next section we shall calculate partition functions for lens spaces \lpq. Since these are closed manifolds, the Turaev-Viro partition function can be viewed as a vacuum-to-vacuum amplitude induced by a manifold with nontrivial topology.

In section 3 we shall turn to manifolds with boundaries. We analyse in detail different triangulations of a genus 1 handlebody $\h_1$, which mediates a topology transition $\O \ra T^2$. An example of genus 2 handlebody is given in section 4.

\sect{Lens Spaces}

The lens spaces \lpq\ are 3-manifolds which have genus 1 Heegard splittings. This means that they are constructed from two solid tori (handlebodies) $\h_1$ by gluing them along the common boundary with a homeomorphism $h_{p,q}: T^2 \ra T^2$.

\begin{figure}
\epsffile{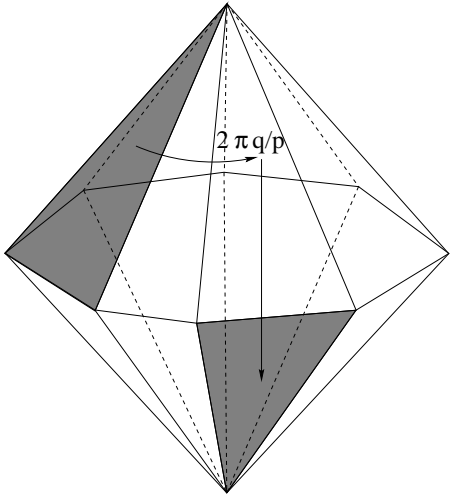}
\caption{Construction of a lens space $L_{p,q}$; the points in the upper half of $\partial \B^3$ are identified with those in the lower one after a rotation with ${2\pi \over p} q$ and a reflection (the shaded faces are glued together).}
\label{lpq}
\end{figure}

Another way of constructing \lpq, more suitable for simplicial manifolds, is the following \cite{rolfsen}. Take the double cone of a $p$-polygon, obtaining thus two solid pyramids glued along the common base which form a 3-ball $\B^3$. Then identify each point in the upper half of $\partial \B^3$ with a point in the lower one after a rotation by ${2\pi \over p} q$ and a reflection in the base plane (see Fig.~\ref{lpq}). The 3-manifold thus obtained is a lens space \lpq, which is defined for integers $p, q$ (relatively prime, $(p,q)=1$), with $q<p$.

Because \lpq\ are prime 3-manifolds \cite{hempel}, they can be interpreted as {\it geons} in the sense of Sorkin \cite{sorkin}, ie topologically nontrivial excitations of the metric.

We calculate the TV partition function $Z(L_{p,q})$ for all lens spaces up to $p=8$. Kauffman and Lins \cite{kl} give the numerical value of these invariants for the case $r=3$. The expresions below are more general, since they are valid for any $r$.

We put $\delta(i,j,k)=1$ if $(i,j,k) \in adm$ and $\delta(i,j,k)=0$ otherwise. All the sums are understood to be over the admissible values of the summands. The first two partition function for $\R \mathrm{P}^3$ and $L_{3,1}$ have also been calculated by Turaev and Viro in their original paper \cite{tv}.

In order to simplify the formulae, we use the notation

\[
\ssj{j_1}{j_2}{j_3}{j_4}{j_5}{j_6}\ \equiv \ \ssj{1}{2}{3}{4}{5}{6}\
\]
and correspondingly, the sums will be $\sum_{i_1} w^2_{i_1} \equiv \sum_1 w^2_1$ etc.

With these conventions, the expressions for the lens spaces we have found are:

\be
Z(L_{2,1} \equiv \R {\mathrm P}^3)= w^{-2} \sum_1 w_1^2
\label{l21}
\ee

\be
Z(L_{3,1})= w^{-2} \sum_1 w_1^2\ \delta(1,1,1)
\label{l31}
\ee

\be
Z(L_{4,1})= w^{-2} \sum_{1,2} w_1^2 w_2^2\ \delta(1,1,2) \ssj{1}{1}{2}{1}{1}{2}
\label{l41}
\ee

\be
Z(L_{5,1})= w^{-2} \sum_{1,2,3} w_1^2 w_2^2 w_3^2\ \ssj{1}{2}{3}{1}{2}{1} \ssj{1}{2}{3}{1}{1}{3} 
\label{l51}
\ee

\be
Z(L_{5,2})= w^{-2} \sum_{1,2} w_1^2 w_2^2\ \ssj{1}{1}{2}{1}{2}{2}
\label{l52}
\ee

\be
Z(L_{6,1})= w^{-2} \sum_{1 \ldots 4} w_1^2 \ldots w_4^2 \ssj{1}{2}{3}{1}{2}{1} \ssj{1}{2}{3}{1}{4}{3} \ssj{1}{3}{4}{1}{1}{4}
\label{l61}
\ee

\be
Z(L_{7,1})= w^{-2} \sum_{1 \ldots 5} w_1^2 \ldots w_5^2 \ssj{1}{4}{5}{1}{1}{5} \ssj{1}{4}{5}{1}{4}{3} \ssj{1}{2}{3}{1}{2}{1} \ssj{1}{2}{3}{1}{4}{3}
\label{l71}
\ee

\be
Z(L_{7,2})= w^{-2} \sum_{1 \ldots 4} w_1^2 \ldots w_4^2 \ssj{2}{3}{2}{4}{1}{1} \ssj{2}{4}{2}{1}{1}{3} \ssj{2}{4}{2}{2}{3}{1}
\label{l72}
\ee

\be
Z(L_{7,3})= w^{-2} \sum_{1,2,3} w_1^2 w_2^2 w_3^2\ \ssj{1}{2}{3}{1}{2}{1} \ssj{1}{2}{3}{1}{3}{3} 
\label{l73}
\ee

\be
Z(L_{8,1})= w^{-2} \sum_{1 \ldots 6} w_1^2 \ldots w_6^2 \ssj{1}{2}{3}{1}{2}{1} \ssj{1}{3}{4}{1}{3}{2} \ssj{1}{4}{5}{1}{4}{3} \ssj{1}{4}{5}{1}{6}{5} \ssj{1}{5}{6}{1}{1}{6}
\label{l81}
\ee

\be
Z(L_{8,3})= w^{-2} \sum_{1,2,3} w_1^2 w_2^2 w_3^2 \ssj{1}{2}{3}{1}{2}{1} \ssj{1}{2}{3}{3}{2}{3}
\label{l83}
\ee

\sect{Handlebodies}

In this section we calculate partition functions for manifolds with boundaries. For a manifold $\M$ with boundary $\partial \M$, the partition function $Z(\M)$ can be interpreted as an amplitude for topology change $\O \ra \partial \M$.

One of the simplest such manifolds is a genus $g$ handlebody $\h_g$ and its boundary is a genus $g$ torus $\partial \h_g \approx T_g$. For $g=0$ this is simple the 3-ball $\B^3$ and some examples of partition function $Z(\B^3)$ have been calculated in \cite{ri}.

A given triangulation of a 3-manifold will be denoted by $(N_0, N_1, N_2, N_3)$, where $N_i$ is the number of $i$-dimensional simplices. A similar notation will be also used for the boundaries (which are 2-dimensional manifolds), i.e. $(N_0,N_1,N_2)$.

The boundary of the genus 1 handlebody $\h_1$ is a torus $T^2$ and all the triangulations of the torus are of the form $(N_0, 3N_0, 2N_0)$ (this follows easily from the Dehn-Sommerville identities). Besides the topology of the manifold, the Turaev-Viro invariant depends also on the colouring of its boundary (and hence of the triangulation of the boundary). In order to distinguish between different triangulations of the same manifold, we attach an extra label representing the triangulation of the boundary, e.g. $\h_1^{(N_0,N_1,N_2)}$.

Although partition functions for the same manifold are different for different triangulations, they have to obey some constraints. Suppose we glue together two copies of a triangulated manifold $\M$ along the common boundary using the identity homeomorphism $Id: \partial \M \ra \partial \M$. The manifold thus constructed is known as the double of the original manifold $2\M$ and it is closed. Therefore the partition function $Z(2\M)$ does not depend of any colouring, since gluing along the boundary implies also summing over all colourings of $\partial \M$. This means we have the following:

\begin{lemma}
Consider two triangulations of an arbitrary manifold $\M$, $\partial \M \ne \O$. Let $(N_0,N_1,N_2)$ and $(P_0,P_1,P_2)$ be their respective boundary triangulations and $j$, $k$ the corresponding boundary coulorings. Then,

\be
\sum_j Z^2(\M^{(N_0,N_1,N_2)}; j) = \sum_k Z^2(\M^{(P_0,P_1,P_2)}; k) = Z(2\M)
\label{2m}
\ee
\end{lemma}

We now particularize the previous lemma for handlebodies. By gluing two copies of a handlebody along the common boundary using the identity homeomorphism and summing over the colouring, we obtain a closed manifold, $S^2 \times S^1$. Since $Z(S^2 \times S^1)=1$ \cite{tv}, the TV partition function for an arbitrarily triangulated 1-handlebody satisfies the identity:

\be
\sum_j Z^2( \h_1^{(N_0,N_1,N_2)}, j) = Z(S^2 \times S^1)= 1
\label{hg1}
\ee

For the 2-handlebody $\h_2$, we have:

\be
\sum_j Z^2( \h_2^{(N_0,N_1,N_2)}, j) = Z[(S^2 \times S^1) \# (S^2 \times S^1)]= Z^{-1}(S^3)
\label{hg2}
\ee
where we used the formula for the connected sum 

\be
Z(A \# B)= \frac{Z(A)\ Z(B)}{Z(S^3)}
\ee

The generalized identity for an arbitrary genus is \cite{ri}:

\be
\sum_j Z^2( \h_g^{(N_0,N_1,N_2)}, j) = [Z(S^3)]^{1-g}= w^{-\chi(T_g)}
\label{hg}
\ee
since $Z(S^3)=w^{-2}$.

Suppose now we take again two copies of a handlebody $\h_1$, but instead of gluing them along the boundary with the identity map, we glue them with the 'twisted' map, ie $h_{0,1}: T^2 \rightarrow T^2$. This maps the meridian of one torus to the longitude of the second. The resulting manifold is a 3-sphere $S^3$. Then the second identity satisfied by partition functions is:

\be
\sum_j Z(\h_1 \cup_{h_{0,1}} \h_1) = Z(S^3)
\ee

The same result can be generalized for an arbitrary handlebody $\h_g$. Gluing two copies along the common boundary with the 'twisted' map $h_{\times}:T_g \rightarrow T_g$, the manifold is again $S^3$. In this case, the map is defined by identifying every meridian on one torus with the corresponding longitude on the second, ie $\mu_i \mapsto \lambda_i,\ i=1..g$.

\be
\sum_j Z(\h_g \cup_{h_{\times}} \h_g) = Z(S^3)
\label{hgx}
\ee

Since the expressions for the partition function become more complicated with the number of boundary edges, the two identities (\ref{hg}) and (\ref{hgx}) provide a method of checking the expressions for $Z(\h_g)$.

\subsection{Genus 1 Handlebodies $\h_1$}

We now construct triangulations for 1-handlebodies with 1, 3, 4 and 9 boundary vertices.

\subsubsection{$\h_1^{(1,3,2)}$}

\begin{figure}
\epsffile{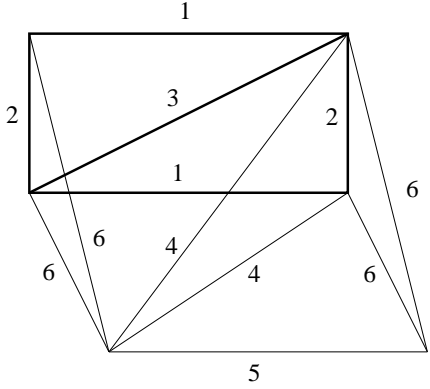}
\caption{ (1,3,2) triangulation of $\h_1$; its $T^2$ boundary is in bold.}
\label{t2lt}
\end{figure}

We start with $\h_1$ triangulated as $(2,6,7,3)$ (see Fig.\ref{t2lt}); its $T^2$ boundary is triangulated as $(1,3,2)$. This triangulation was considered by Louko and Tuckey \cite{louko} in the context of Regge calculus applied to quantum cosmology. The partition function is:

\be
Z(\h_1^{(1,3,2)}, j)= w^{-3} w_1 w_2 w_3\ \sum_{4,5,6} w_4^2 w_5^2 w_6^2 \ssj{1}{2}{3}{6}{4}{6} \ssj{1}{2}{3}{4}{6}{4} \ssj{2}{4}{4}{5}{6}{6}
\label{zh1}
\ee
where $j \equiv (1,2,3)$ denotes the collective colouring of the $T^2$ boundary.

\subsubsection{$\h_1^{(3,9,6)}$}

\begin{figure}
\epsffile{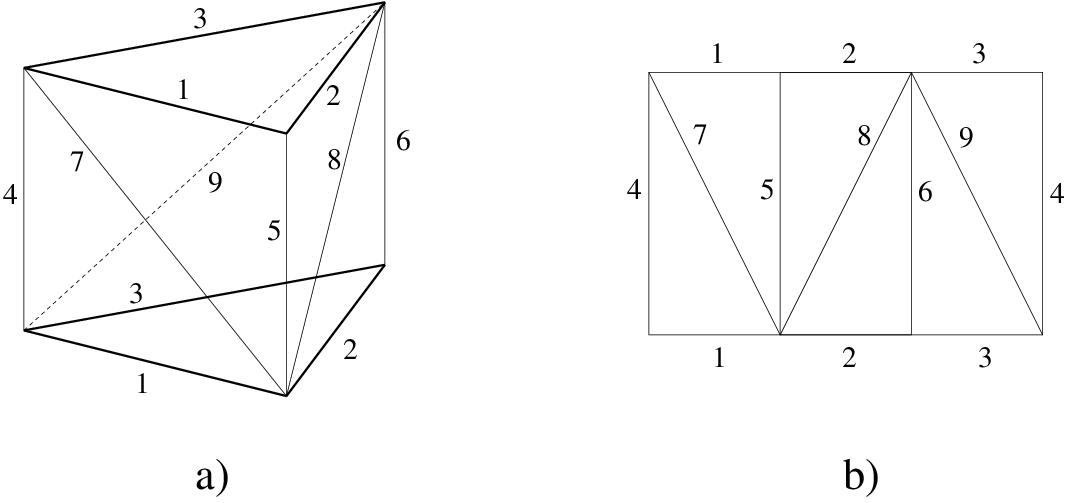}
\caption{(a) (3,9,6) triangulation of $\h_1$, constructed from a prism with two opposite faces (in bold) identified; (b) its boundary, $T^2$.}
\label{h3}
\end{figure}

This can be obtained from a triangular prism after gluing the top and the bottom bases (see Fig.~\ref{h3}).

\be
Z(\h_1^{(3,9,6)}, j)= w^{-3} w_1 \ldots w_9 \ssj{1}{2}{3}{8}{7}{5} 
\ssj{1}{2}{3}{6}{9}{8} \ssj{1}{8}{9}{3}{4}{7}
\label{zh3}
\ee

\subsubsection{$\h_1^{(4,12,8)}$}

\begin{figure}
\epsffile{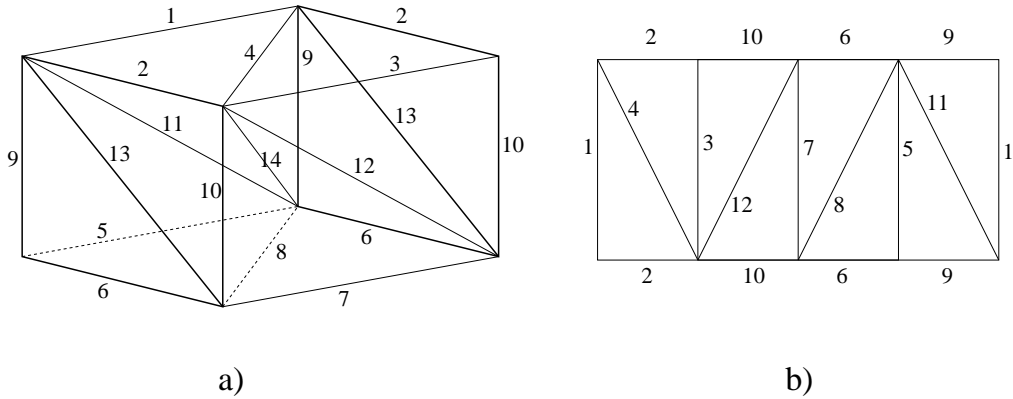}
\caption{(a) (4,12,8) triangulation of $\h_1$, constructed from a cube with two opposite faces (in bold) identified; (b) its boundary, $T^2$.}
\label{h4}
\end{figure}

Starting from a solid cube, we identify the front and the back face, forming thus a solid torus $\h_1$, as in Fig.~\ref{h4}. Its $T^2$ boundary is triangulated as $(4,12,8)$.

We have 4 vertices, 14 edges and 6 tetrahedra: $a=e=4$, $b=14$, $f=12$, $d=6$. The partition function is:

\[
Z(\h_1^{(4,12,8)},j)=w^{-4}\ w_1 \ldots w_{12} \sum_{13,14} w^2_{13} w^2_{14} 
\ssj{1}{2}{4}{14}{9}{11} \ssj{5}{6}{8}{13}{11}{9} \cdot \]

\be
\ssj{11}{13}{8}{10}{14}{2}
\ssj{6}{7}{8}{10}{14}{12}
\ssj{2}{3}{4}{12}{13}{10}
\ssj{12}{14}{6}{9}{13}{4}
\label{zh4}
\ee
where $j \equiv (1, \ldots 12)$ is again the collective boundary colouring.

\subsubsection{$\h_1^{(9,27,18)}$}

\begin{figure}
\epsffile{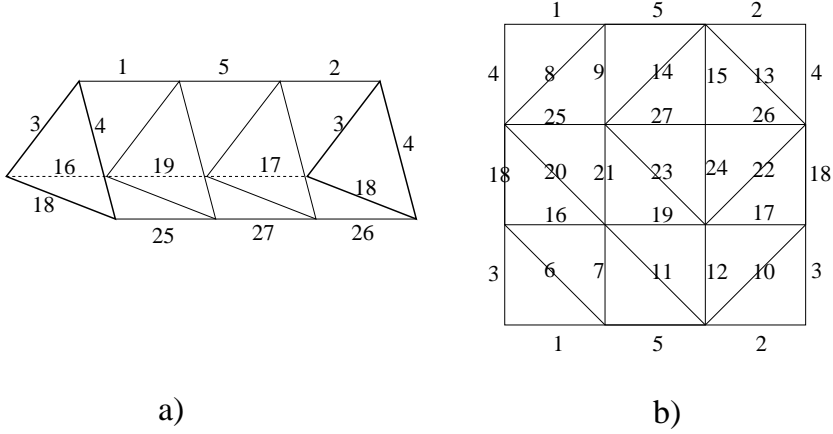}
\caption{(a) (9,27,18) triangulation of $\h_1$, constructed from a long triangular prism with two opposite faces (in bold) identified; (b) its boundary, $T^2$.}
\label{h9}
\end{figure}

A triangulation with 9 boundary vertices can be obtained from 3 prisms glued one on top of the other as in Fig.~\ref{h9}. We have for the partition fuction:

\[
Z(\h_1^{(9,27,18)},j)=w^{-9}\ w_1 \ldots w_{27} \ssj{3}{18}{4}{8}{1}{6} \ssj{6}{18}{8}{20}{7}{16} 
\ssj{7}{21}{9}{25}{8}{20}  \cdot \]

\[
\ssj{7}{21}{9}{14}{5}{11} \ssj{11}{21}{14}{23}{12}{19} 
\ssj{12}{24}{15}{27}{14}{23} \ssj{3}{18}{4}{13}{2}{10} \cdot \]

\be
\ssj{10}{18}{13}{22}{12}{17} \ssj{12}{24}{15}{26}{13}{22}
\label{zh9}
\ee

From these examples, it is easy to see how we can construct 1-handlebodies with arbitrary boundary triangulations. Tedious but straightforward calculations show that all the partition functions (\ref{zh1}) -- (\ref{zh9}) verify the identity (\ref{hg1}). Since the (1,3,6) and (9,27,18) triangulations have symmetric $T^2$ boundaries (both are squares), we can also check the identity (\ref{hgx}) by gluing the two $\h_1$'s with the 'twisted' map. In both cases we recover the known partition function $Z(S^3)$.

\sect{A genus 2 handlebody $\h_2$}

We now turn to the genus 2 handlebody $\h_2$. We can obtain $\h_2$ by gluing with corners two $\h_1$'s along a disc embedded in the common boundary. For this, we need the following prescription.

\subsection{Gluing with corners: prescription for summation}

Suppose we glue two manifolds $\M_1$ and $\M_2$ not on an entire component of the boundary, but only on a surface which is strictly included in one of such components $\Sigma \subset \partial \M_1$ (the so-called {\it gluing with corners}).
We assume that $\Sigma$ has $l$ edges, from which the first $k$ are on $\partial \Sigma$.

In this case the partition function of the resulting manifold $\M_1 \cup_\Sigma \M_2$ will be 

\be
Z(\M_1 \cup_\Sigma \M_2; i_1 .. i_k, p, p')= \sum_{i_{k+1} \ldots i_l} Z(\M_1, i, p) Z(\M_2, i, p') \frac{w^{\#\ vertices\ of\ \partial \Sigma}}{ w_{i_1} \ldots w_{i_k} }
\ee
where $i\equiv (i_1, \ldots ,i_k, i_{k+1}, \ldots i_l)$ are the edges of $\Sigma$ and the labels $p$, $p'$ denotes the edges in $\partial M_1 \setminus \Sigma$ and $\partial M_2 \setminus \Sigma$, respectively.

Thus, the prescription for gluing with corners is that we should multiply by a factor

\be
\frac{w^{\#\ vertices\ of\ \partial \Sigma}}{ w_{i_1} \ldots w_{i_k} }
\ee
where $i_1 \ldots i_k$ are the edges of $\partial \Sigma$ and therefore they are still on $\partial \M$ after gluing.

If the gluing is done on a whole component of $\partial \M$, there are no vertices and edges left on that component and thus the previous factor is 1, recovering the usual expression for gluing.

\subsection{$\h_2^{(14,48,32)}$}

Having the previous prescription for summation, it it easy to construct a genus 2 handlebody.

We start with two 1-handlebodies $\h_1^{(9,27,18)}$ as in Fig.~\ref{h9}; the second handlebody is the mirror image of the first one and will have prime indices. We glue them along the disk formed by the two triangles labelled $5,7,11,12,19$. Taking into account the previous prescription, we obtain the following expression:

\[
Z(\h_2^{(14,48,32)})= w^{-14} w_1 \ldots w_{10} w_{12} \ldots w_{27} w_{1'} \ldots w_{4'} w_{6'} w_{8'} w_{9'}
w_{10'} w_{13'} \ldots w_{18'} \cdot \]

\[ w_{20'} \ldots w_{27'}
\ssj{3}{18}{4}{8}{1}{6} \ssj{6}{18}{8}{20}{7}{16} \ssj{7}{21}{9}{25}{8}{20} \ssj{12}{24}{15}{27}{14}{23}
 \cdot
\]

\[
\ssj{3}{18}{4}{13}{2}{10} \ssj{10}{18}{13}{22}{12}{17} \ssj{12}{24}{15}{26}{13}{22}
\ssj{3'}{18'}{4'}{8'}{1'}{6'}\ssj{6'}{18'}{8'}{20'}{7}{16'}  \cdot
\]

\[
\ssj{7}{21'}{9'}{25'}{8'}{20'} \ssj{12}{24'}{15'}{27'}{14'}{23'} \ssj{3'}{18'}{4'}{13'}{2'}{10'}
\ssj{10'}{18'}{13'}{22'}{12}{17'} \ssj{12}{24'}{15'}{26'}{13'}{22'} \cdot
\]

\[
\sum_{11} w^2_{11} \ssj{7}{21}{9}{14}{5}{11} \ssj{11}{21}{14}{23}{12}{19} \ssj{7}{21'}{9'}{14'}{5}{11}
\ssj{11}{21'}{14'}{23'}{12}{19}
\]

Again, it can be checked that the above partition function satisfies eq.~(\ref{hg2}).

\sect{Conclusions}

In order to explore the consequences of interpreting the Turaev-Viro state sum model as a theory of quantum gravity in three dimensions, we need to be able to evaluate that state sum for the whole range of closed three manifolds to find the partition function. Transition amplitudes between 2-manifolds with different topologies can similarly be calculated from the Turaev-Viro expression for manifolds with boundaries.

In this paper we have evaluated the TV invariant for a number of closed manifolds, the lens spaces \lpq\ for values of $p$ up to 8 and general $q$. We have also calculated the state sum for genus 1 and 2 handlebodies, the results of which are useful firstly as transition amplitudes for manifolds with boundaries, and secondly as factors in any general project to calculate the TV invariant for closed manifolds by building them from handlebodies. Future work involves seeing precisely how this can work and calculating the state sum for a handlebody of arbitrary genus.

~\\
{\Large\bf Acknowledgements}\\
~~\\
We thank Jane Paterson for useful discussions on 3-manifolds topology. One of the authors (R.I.) has been kindly supported by Cambridge Overseas Trust, the Ra\c tiu Foundation and ORS.

\end{document}